
\def\twelvepoint{
  \font\twelverm=cmr12   \font\twelvei=cmmi12	
  \font\twelvebf=cmbx12  \font\twelvesy=cmsy10 scaled\magstep 1
  \font\twelveit=cmti12  \font\twelvett=cmtt12  \font\twelvesl=cmsl12
  \font\tenrm=cmr10    \font\teni=cmmi10	
  \font\tensy=cmsy10   \font\tenbf=cmbx10
  \font\ninerm=cmr9    \font\ninei=cmmi9	
  \font\ninesy=cmsy9   \font\ninebf=cmbx9
  \font\eightrm=cmr8   \font\eighti=cmmi8	
  \font\eightsy=cmsy8  \font\eightbf=cmbx8
  \font\fiverm=cmr5    \font\fivei=cmmi5	
  \font\fivesy=cmsy5   \font\fivebf=cmbx5
  \textfont0=\twelverm   \scriptfont0=\eightrm  \scriptscriptfont0=\fiverm
  \textfont1=\twelvei    \scriptfont1=\eighti   \scriptscriptfont1=\fivei
  \textfont2=\twelvesy   \scriptfont2=\eightsy  \scriptscriptfont2=\fivesy
  \textfont3=\tenex      \scriptfont3=\tenex    \scriptscriptfont3=\tenex
  \def\rm{\fam0\twelverm}  \def\mit{\fam1}  \def\cal{\fam2}
  \textfont\itfam=\twelveit  \def\it{\fam\itfam\twelveit}
  \textfont\slfam=\twelvesl  \def\sl{\fam\slfam\twelvesl}
  \textfont\ttfam=\twelvett  \def\tt{\fam\ttfam\twelvett}
  \textfont\bffam=\twelvebf  \scriptfont\bffam=\eightbf
  \scriptscriptfont\bffam=\fivebf  \def\bf{\fam\bffam\twelvebf}
  \let\sc=\tenrm  \let\sb=\ninebf
  \setbox\strutbox=\hbox{\vrule height8.5pt depth3.5pt width0pt}
  \def\doublespace{\baselineskip=24pt   \lineskip=0pt \lineskiplimit=-5pt}
  \def\singlespace{\baselineskip=13.5pt \lineskip=0pt \lineskiplimit=-5pt}
  \def\oneandahalf{\baselineskip=18pt   \lineskip=0pt \lineskiplimit=-5pt}
  \parindent=.5in  \parskip=7pt plus 1pt minus 1pt
  \footline={\ifnum\pageno=1\hfil\else\hss\twelverm-- \folio\ --\hss\fi}
  \def\simlt{\mathrel{\hbox to 0pt{\lower 3.5pt\hbox{$\mathchar"218$}\hss}
      \raise 1.5pt\hbox{$\mathchar"13C$}}}
  \def\simgt{\mathrel{\hbox to 0pt{\lower 3.5pt\hbox{$\mathchar"218$}\hss}
      \raise 1.5pt\hbox{$\mathchar"13E$}}}
  \hsize=170mm\vsize=230mm\rm\raggedbottom\doublespace
}

\def\caidx{Ca{\sc II}~H + H$\epsilon$/Ca{\sc II}~K\ }
\def\b4{$B_{\nu4000}$}
\def\mz{$\langle z\rangle$\ }

\def\centsh#1{
\vskip24pt
\centerline{#1}

\nobreak}

\def\centdbl#1#2{
\vskip24pt
\centerline{#1}
\nobreak
\centerline{#2}

\nobreak}


\def\Rf{\par\parskip=0pt plus 1pt\noindent\hangindent0.4in\hangafter1}

\twelvepoint
\voffset=-.3cm
\hoffset=0in
\vsize=9in
\hsize=6.37in
\baselineskip=25pt
\raggedbottom
\vskip 4.0in
\null\vskip.5in
\centerline{\bf EVOLUTION OF GALAXIES IN CLUSTERS}
\vskip.4in
{\baselineskip=13.5pt
\centerline{
S{\sc T\'EPHANE}\ C{\sc HARLOT}$^{1}$\footnote{}{\vskip-19pt \hskip-1.5cm
$^1$ Also affiliated with the Center for Particle Astrophysics, University of
California at Berkeley and the Institute of Geophysics and Planetary Physics,
Lawrence Livermore National Laboratory}
and J{\sc OSEPH} S{\sc ILK}$^{2}$\footnote{}{\hskip-1.5cm
$^2$ Also affiliated with the Department of Physics and the Center for Particle
Astrophysics, University of California at Berkeley}
}
\par}
\centerline{Department of Astronomy, University of California, Berkeley, CA
94720}
\vskip2in
\centerline{Accepted for publication in the {\it Astrophysical Journal}}

\vfill\eject

\centerline {\bf ABSTRACT}
We use new models of stellar population synthesis to estimate the fraction of
stars formed during the last major bursts of star formation in E/S0 galaxies in
low-redshift clusters ($z\simlt0.4$) from the spectral signatures of
intermediate-age stars. We find that the mass fraction of stars formed in late
bursts in early-type galaxies in clusters must have decreased smoothly with
redshift, from about 8\% at $z\approx1$ to less than 1\% at $z\approx0$. This
result, which we interpret as a constraint on stellar mass added in mergers, is
nearly independent of the assumed ages and morphological types of the
progenitor galaxies prior to the last major bursts of star formation. We then
compute the implied color and 4000 {\AA} break evolution for progenitors of
E/S0 galaxies in clusters at redshifts $0\simlt z\simlt1$. We investigate a
conservative model in which all present-day E/S0 galaxies are assumed to
initially be elliptical galaxies and to then undergo bursts of star formation
at the rate estimated from the signatures of intermediate-age stars at low
redshifts. This model reproduces well the observed spread of colors and 4000
{\AA} breaks of galaxies in high-redshift clusters, but underestimates the
fraction of galaxies in the blue tail of the distribution. Such a discrepancy
may be interpreted as an increasing fraction of spiral galaxies in clusters at
high redshift, as suggested independently by recent {\it HST} observations of
intermediate-redshift clusters. The current data do not seem to require
morphological evolution of cluster galaxies out to $z\sim0.4$, but suggest that
either morphological or strong luminosity evolution might have played a major
role at $z\simgt0.7$. Our results provide spectrophotometric corroboration to
high redshift of a common, but poorly justified, interpretation of the
Butcher-Oemler effect.

\vskip1.7cm
\noindent
{\it Subject Headings:} cosmology --- galaxies: clusters --- galaxies:
evolution --- galaxies: formation --- galaxies: interactions --- galaxies:
stellar content

\vfill \eject

\centsh{\bf 1. INTRODUCTION}

Clusters are valuable laboratories for studying galaxy evolution over large
look-back times because they are more readily identifiable than field galaxies
at high redshifts. The extent to which galaxy evolution differs between cluster
and field environments is not clear. Present-day clusters are rich in
elliptical galaxies, whereas the field population is rich in spiral galaxies.
Many galaxies in distant clusters have red colors consistent with the
hypothesis that these galaxies formed coevally at a redshift $z>2$ and evolved
passively to become the E/S0 galaxies which dominate present-day clusters such
as Coma (e.g., Bower, Lucey, \& Ellis 1992; Arag\'on-Salamanca et al. 1993).
However, there is also strong evidence for the evolution of galaxies in
clusters. Butcher \& Oemler (1978, 1984) pointed out that the fraction of blue
galaxies in the cores of rich clusters increases with redshift, from about 2\%
at $z \approx0$ to nearly 20\% at $z\approx0.4$ --- the ``Butcher-Oemler"
effect (see also Dressler \& Gunn 1983, 1992; Couch \& Sharples 1987; Lavery \&
Henry 1988). These blue galaxies are generally interpreted as star-forming
galaxies which have disappeared today, either because they merged or because
they were only temporarily blued by interaction-induced starbursts (Dressler \&
Gunn 1983, Lavery \& Henry 1988; Evrard 1991). Recently, {\it HST} observations
have revealed that, in reality, many blue galaxies in several clusters at
$z\approx0.4$ are normal spiral or at least ``disky" galaxies, which do not
show signs of interactions and vigorous bursts of star formation (Couch et al.
1993; Dressler 1993).

The evidence for evolution of cluster galaxies also comes from absorption-line
studies of E/S0 galaxies in low-redshift clusters ($z\simlt0.4$). Many of the
galaxies which were observed exhibit spectral signatures consistent with the
presence of intermediate-age stellar populations (Couch \& Sharples 1987;
Lavery \& Henry 1988; Pickles 1989; Dressler \& Gunn 1992). This would indicate
that the galaxies were forming stars, and hence were bluer, only a few billion
years ago. However, the fact that many blue galaxies in distant clusters appear
to be normal spiral galaxies suggests that morphological evolution of spiral
into E/S0 galaxies could also have played a significant role over the last few
billion years. The situation is complicated by the fact that present-day
clusters contain a significant fraction of gas-poor, spiral galaxies with
colors nearly as red as E/S0 galaxies (with $B-V\approx0.9$; see the recent
review by Oemler 1992). On average, about 10$-$20\% of galaxies in the cores of
nearby rich, concentrated clusters are of late type (Dressler 1980; Butcher \&
Oemler 1978), whereas only 2$-$5\% are blue (where a blue galaxy is defined as
having a $B-V$ color at least 0.2 mag bluer than the E/S0 sequence at the same
absolute magnitude; Butcher \& Oemler 1984). In this context, the
absorption-line signatures of intermediate-age stars in E/S0 galaxies represent
an invaluable means of estimating the mininum fraction of blue galaxies in
high-redshift clusters which could have become E/S0 galaxies today, regardless
of morphological evolution.

In this paper we use stellar population synthesis models to estimate the
implications of the signatures of intermediate-age stars in E/S0 galaxies in
low-redshift clusters for the photometric evolution of these galaxies at high
redshifts, when the stars formed. Our main purpose is to evaluate the minimum
color evolution implied by the low-redshift spectral analyses, and hence to
infer the additional evolution (of colors or morphology) needed to explain the
blue galaxy content of high-redshift clusters. We first describe in \S2 the
signatures of the past history of star formation in the spectra of passively
evolving galaxies using new models of stellar population synthesis. We use
these arguments in \S3 to interpret the results of absorption-line studies of
E/S0 galaxies in low-redshift clusters ($z\simlt0.4$) in terms of the average
rate of star formation as a function of redshift in progenitors of E/S0
galaxies ($z\simlt1$). We then compute in \S4 the implied spectrophotometric
evolution for progenitors of E/S0 galaxies at high redshift, and compare this
with direct observations of galaxies in distant clusters. We discuss in \S5 the
implications of our results for the color and morphological evolution of
galaxies in clusters. Most of the arguments presented throughout this paper
pertain to the evolution of galaxies in the cores of rich, concentrated
clusters.

\centsh{\bf 2. PREDICTIONS FROM POPULATION SYNTHESIS MODELS}

We first describe the signatures of the past history of star formation in the
spectra of passively evolving galaxies using the Bruzual \& Charlot (1993)
models of stellar population synthesis. The models have solar metallicity and
include all phases of stellar evolution from zero-age main sequence to
supernova explosion (for progenitors more massive than $8\,M_{\odot}$) or the
end of the white-dwarf cooling sequence (for less massive progenitors). Most of
the stellar spectra were taken directly from observations of Galactic stars at
near-ultraviolet, optical, and near-infrared wavelengths and were extended into
the far ultraviolet using model atmospheres. The main adjustable parameters in
the population synthesis models are the star formation rate and the initial
mass function (IMF). Here, we are primarily interested in the evolution of
galaxies which stop forming stars after a last burst of star formation. We
consider two extreme histories of star formation for the underlying galaxy: a
burst of duration $5\times10^8$ yr (model elliptical galaxy) and a constant
star formation rate (model spiral galaxy). In each case, we assume that star
formation stops when the galaxy undergoes a last burst of star formation, of
duration $10^8$ yr. The age of the underlying elliptical or spiral galaxy at
this time and the mass fraction of the burst --- defined throughout this paper
as the fraction of the final mass in stars of the galaxy which is formed during
the burst --- are kept as free parameters. We further adopt in all models the
Salpeter IMF with lower and upper cutoffs of $0.1$ and $100\,M_{\odot}$,
respectively.

Figure 1 shows the evolution of the optical/infrared colors and 4000 {\AA}
break of the model spiral and elliptical galaxies when the burst of star
formation occurs at an age of 6 Gyr and involves 10\% of the total mass (we
have adopted here the definition of the 4000 {\AA} break given by Bruzual 1983,
i.e., the dimensionless ratio of the fluxes in two bands just redward and
blueward of the 4000~{\AA} discontinuity). Before the burst, the colors of the
model elliptical galaxy redden rapidly as the most massive stars leave the main
sequence, and the 4000 {\AA} break becomes strong (the red phase around 1 Gyr,
which is mostly visible in the $I-K$ color, is produced when stars on the
asymptotic giant branch dominate the spectrum; see Bruzual \& Charlot 1993).
Meanwhile, the colors and 4000 {\AA} break of the model spiral galaxy remain
almost constant, since they are dominated by massive stars that are
continuously replenished on the main sequence. During the burst, at an age of 6
Gyr, the new input of massive stars blues the colors and reduces the 4000 {\AA}
break temporarily in both models. From then on, the evolution of the colors and
4000 {\AA} break depend only weakly on the past history of star formation. The
difference between the two models in Figure 1 is less than 0.1 mag in all
colors and less than 10\% in the 4000 {\AA} break at all ages after 6 Gyr.
Moreover, the colors and 4000 {\AA} break at ages greater than 6.5 Gyr are
nearly the same as those of the model elliptical galaxy before the burst.
Hence, we conclude from Figure 1 that the age and history of star formation of
a galaxy which has evolved passively for several billion years cannot be
determined from only the observed continuum emission. The situation is even
further complicated when corrections for reddening have to be made.

Fortunately, more precise information on the past history of star formation in
passively evolving galaxies can be learned from the absorption lines of
hydrogen and of other prominent atoms and molecules such as Mg, Mg$_2$, Fe, Ca,
Na, Sr, and CN (O'Connell 1980, 1986; Burstein et al. 1984; Rose 1985; Couch \&
Sharples 1987; Pickles 1985; Worthey, Faber, \& Gonzalez 1992; Jablonka \&
Alloin 1993). We now illustrate this. The medium wavelength resolution and
unique metallicity of the models used here restrict our analysis to only a few
absorption lines. Two suitable examples are the equivalent width of the
H${\delta}$ absorption line and the ratio of the central intensity of the
combined Ca{\sc II} H and H$\epsilon$ lines to the central intensity of the
Ca{\sc II} line, referred to as the ``\caidx index" (Rose 1985). These features
both hardly depend on metallicity (Rose 1984, 1985). The \caidx index, which is
constant in F, G, and K stars, drops significantly in A and B stars as the
Ca{\sc II} lines weaken and the H-Balmer lines strengthen. Therefore, a strong
H$\delta$ absorption equivalent width indicates the presence of stars younger
than about 2 Gyr, and a strong \caidx index that of older, longer-lived stars.

We show in Figure 2 the evolution of the H$\delta$ absorption equivalent width
and the \caidx index as a function of $U-V$ and $V-K$ colors for the two models
of Figure 1. The evolution is shown starting from the age of 6 Gyr, when the
burst occurs ({\it triangles}). At this age, the model spiral galaxy has a
stronger H$\delta$ equivalent width and a weaker \caidx index than the model
elliptical galaxy because it contains some A and B stars. Shortly after the
burst, at an age of 6.5 Gyr, the $U-V$ and $V-K$ colors of both models already
approach those characteristic of passively evolving galaxies (see Fig.~1).
However, the H$\delta$ equivalent width and \caidx index continue to evolve
significantly for about 1.5 Gyr, with a weak dependence on the past history of
star formation in the underlying galaxy. In particular, from 6.5 to 8 Gyr, the
H$\delta$ absorption equivalent width of the model galaxies  drops by an order
of magnitude and the \caidx jumps by more than 50\%, while the colors redden by
less than 0.1 mag. Therefore, stellar absorption lines can reveal recent bursts
of star formation in galaxies with colors otherwise typical of old, passively
evolving stellar populations. We have considered here only the cases of the
H$\delta$ equivalent width and \caidx index, whose sensitivities are limited to
roughly 2 Gyr after a burst of star formation. However, other combinations of
prominent metallic features can help detect bursts of star formation which are
several billion years old (e.g., Pickles 1985).

Absorption-line strengths can also be used to untangle the competing effects of
metallicity and age on the spectra of passively-evolving galaxies. Stars of
increasing metallicity have redder colors and evolve more slowly (e.g.,
Schaller et al. 1992). Thus, the age assigned to a stellar population from its
observed colors is a strong function of metallicity. Pickles (1985; 1989; see
also Pickles \& van de Kruit 1988) has investigated the effects of age and
metallicity on galaxy spectra by using isochrone spectra of instantaneous-burst
stellar populations for six different ages between 2 and 15 Gyr and four
metallicities between 0.0017 and 0.06. He showed that tight constraints on the
age and metallicity of the various generations of stars in a galaxy can be
obtained by requiring the observed spectrum to be fitted with a combination of
isochrone spectra which reproduces the prominent metallic lines in addition to
the continuum emission. One then infers the fraction of light, generally in the
optical region of the spectrum, produced by stars of a given age.
Unfortunately, we cannot illustrate this method here because our population
synthesis models are restricted to solar metallicity. We refer the reader to
the studies by Pickles (1985), Jablonka \& Alloin (1993), and Worthey (1993).

We also emphasize here that the contribution to the optical light by the
youngest stars in a galaxy provides valuable information on the fraction of the
galaxy mass involved in the last major burst of star formation. Figure 3 shows
the evolution of the fractional contribution to the visual luminosity of
various model galaxies by stars created in a last burst. As expected, the
fraction of $V$ light produced by new stars in a model elliptical galaxy scales
with the fraction of the galaxy mass involved in the last burst of star
formation (Fig.~3{\it a}). The contribution peaks during the burst, when new
massive stars occupy the main sequence, and then drops to a nearly constant
value. For example, when 2\% and 10\% of the stars are created during the
burst, their contribution to the V light 2 Gyr later is 5\% and 24\%,
respectively. The reason for the nearly constant contribution at late ages is
that the visual luminosity of a passively evolving stellar population changes
only slowly after a few billion years (see Fig.~1 of Bruzual \& Charlot 1993).
In fact, this also implies that the fraction of $V$ light produced by young
stars depends only weakly on the age of the underlying population at the time
of the burst, as Figure 3{\it b} shows. Figure 3{\it c} further indicates that
the predicted contribution by stars formed in a last burst to the visual
luminosity is only about 20\% lower when the underlying galaxy is a model
spiral instead of elliptical galaxy. Hence, the contribution by young stars to
the visual luminosity of a passively evolving galaxy can be used to estimate
the fraction of mass involved in the last burst of star formation, almost
independently of the age and past history of star formation of the underlying
galaxy when the burst occured.

\centsh{\bf 3. RATE OF RECENT STAR FORMATION IN EARLY-TYPE GALAXIES}

In this section, we estimate the rate of recent star formation in early-type
galaxies at low redshifts ($z\simlt0.4$) from the absorption-line signatures of
intermediate-age stars, using the models and arguments presented in the
previous section. Since we cannot directly perform spectral analyses based on
many spectral lines and include the effects of varying metallicity, we use here
the results of several detailed absorption-line studies of low-redshift E/S0
galaxies that were published over the last few years. These results are
conveniently expressed in terms of the fraction of the optical light of the
galaxies accounted for by stars formed in the last few billion years. From the
fractional contributions to the optical light, we can then securely estimate
the mass of stars formed (\S2). The galaxies studied at low redshifts were
mostly selected from the field, but the few examples in clusters present
similar characteristics (O'Connell 1980; Burstein et al. 1984; Rose 1985;
Gonzalez 1993). At higher redshifts, spectral analyses have mainly been
performed on cluster galaxies (Dressler \& Gunn 1982, 1983, 1992; Couch \&
Sharples 1987; Lavery \& Henry 1988; Mellier et al. 1988; Pickles 1989;
Jablonka \& Alloin 1993). Most of the galaxies in question have luminosities
around $L_\star$, which correspond to masses in the range $0.5-1.5\times
10^{11} h^{-2} M_\odot$. We do not consider here the first-ranked, giant cD
galaxies observed in some rich clusters, which have typical luminosities
$\simgt10L_{ \star}$ and are believed to accrete up to 40\%$-$75\% of their
mass in a Hubble time by cannibalizing other cluster galaxies (e.g., Lauer
1988). We now interpret these observations.

Rose (1985) has analyzed the absorption lines of a sample of 12 nearby E/S0
galaxies with luminosities in the range $0.01L_{\star}\simlt L\simlt
L_{\star}$. He found that the \caidx indices of the galaxies have a sharp
distribution (except for one outlier, NGC~205) around a mean value
corresponding to the index of the prototypical elliptical galaxy
M32.$^{3}$\footnote{}{\vskip-19pt $^3$ The weak dependence of the \caidx index
on luminosity does not contradict the increasing strength of prominent metallic
lines (Mg$_2$, Fe, etc.) of E/S0 galaxies with luminosity. In fact, the latter
is generally attributed to a change in metallicity (e.g., Worthey, Faber, \&
Gonzalez 1993). The \caidx index, on the other hand, depends only weakly on
metallicity (Rose 1985).} This dwarf galaxy is one of the best studied nearby
E/S0 galaxies, and it exhibits several signatures of recent star formation
(Spinrad \& Taylor 1969; Faber 1972; Pritchet 1977; O'Connell 1980, 1986;
Burstein et al. 1984; Rose 1985; Freedman 1992). The common value of the \caidx
index in nearly all the galaxies studied is significantly lower than the value
expected from the proportions of main-sequence dwarfs and red giants indicated
by the Sr~{\sc II}/Fe~{\sc I} indices. Rose concludes that the discrepancy
between the \caidx and Sr{\sc II}~/Fe~{\sc I} indices implies that roughly 2\%
of the light at 4000 {\AA} in the galaxies is produced by extra populations of
A and B stars. Since the evolutionary phase of the stars is not known, this
sets an upper limit on the contribution by main-sequence A and B stars with
lifetimes less than 2.5 Gyr. Using diagrams similar to those shown in Figure~2,
we infer that typically less than 0.5\% of the mass in stars in nearby E/S0
galaxies have formed within the last 2.5 Gyr.

The observed color dispersion of nearby E/S0 galaxies provides another upper
limit on their young star content, although this requires accurate photometry
(see Fig.~1). Early-type galaxies follow a color-magnitude relation, which is
generally attributed to an increase in metallicity from faint to bright
luminosities (e.g., Faber 1977; Bower, Lucey, \& Ellis 1992). Schweizer \&
Seitzer (1992) show that the departure of the $U-B$ and $B-V$ colors of nearby
E/S0 galaxies from the mean color-magnitude relation correlates with their
``fine-structure index" (a measure of dynamical youth) and absorption-line
strengths. Thus, the dispersion of E/S0 galaxies around the mean
color-magnitude relation could originate, at least in part, from an age
difference. Schweizer \& Seitzer consider a model in which they assume that
E/S0 galaxies result from equal-mass mergers of two spiral galaxies of the same
type (Sb-Sb or Sc-Sc). They adopt three main parameters: the star formation
timescale, $\tau$, of the model spiral galaxies (6 Gyr for Sb galaxies and 10
Gyr for Sc galaxies); the mass fraction $\epsilon$ of the available gas
converted into stars at the time of the merger (typically, 10\%); and the age
of the spiral galaxies (15 Gyr). Schweizer \& Seitzer then use population
synthesis models to interpret the relative blueing of each of the 69 galaxies
of their sample in terms of the time $t_{\rm m}$ elapsed since the merger. The
results range from $4\simlt t_{\rm m} \simlt10$ Gyr for Sb-Sb mergers to
$5\simlt t_{\rm m} \simlt11$ Gyr for Sc-Sc mergers.  The galaxy mass fraction
converted into stars during the merger can then be expressed as $\epsilon
\exp[(t_{\rm m}-15) /\tau]$, where $t_{\rm m}$ and $\tau$ are in Gyr.

E/S0 galaxies at higher redshifts also show many signs of recent star
formation. Dressler \& Gunn (1983, 1992), Lavery \& Henry (1988), and Jablonka
\& Alloin (1993) have discovered many red galaxies in clusters at $0.3\simlt
z\simlt0.6$ with strong Balmer absorption lines characteristic of
post-starburst spectra. Pickles (1989, see \S2 of the present paper) has
spectroscopically evaluated the ages of young stellar populations and their
contribution to the visual luminosity in bright ($R\simlt20$~mag), red
($B-R\simgt2.0$~mag) galaxies in clusters at redshifts between 0.18 and 0.39
(with absolute luminosities $0.4L_{\star}\simlt L\simlt 2.5 L_{\star}$). To
increase the signal-to-noise ratio in the observations, he coadded the spectra
of the galaxies observed in each of the six clusters in his sample,
distinguishing between a faint ($\simlt L_{\star}$) and a bright ($\simgt
L_{\star}$) class. This procedure is justified by the small differences in
colors and line strengths of individual spectra in a given class. In fact, the
inferred fraction of $V$ light accounted for by young (2$-$6~Gyr) stars appears
to be roughly comparable for the two luminosity classes in each cluster.
Therefore, we adopt here the average over faint and bright galaxies of the
fraction of visual luminosity contributed by young stars as a measure of the
rate of recent star formation in E/S0 galaxies in each cluster. This fraction
appears to increase with redshift, from about 7\% at $z=0.18$ to about 12\% at
$z=0.39$. The population synthesis models indicate that the corresponding
fraction of the galaxy masses transformed into stars $2-6$~Gyr earlier ranges
from roughly 4\% at $z=0.18$ to 8\% at $z=0.39$.

Couch \& Sharples (1987) have also analyzed the stellar content of bright
($R\simlt20$~mag), red ($B_J-R\simgt2.0$~mag) galaxies in three rich clusters
at $z=0.31$. Unlike Pickles (1989), they identified marked differences in the
signatures of recent star formation among the galaxies, which they classified
into ``normal" E/S0 galaxies (62 objects) and E/S0 galaxies with strong
H$\delta$ absorption equivalent widths (11 objects with EW[H$ \delta$]$ \geq3
$~{\AA}). Couch \& Sharples measured the \caidx indices of the coadded spectra
of the two types of galaxies, but did not determine the relative contributions
to the light by main-sequence dwarfs and red giants. An upper limit on the
number of young A and B stars, however, may be obtained from the difference
between the observed indices and the value corresponding to the case of a red
giant-dominated stellar population. The result is that less than 10\% of the
light at 4000~{\AA} in the normal E/S0 galaxies is produced by A and B stars.
This implies that less than 1.8\% of the stars formed in the last 2.5 Gyr. The
coadded spectrum of the strong-lined galaxies, on the other hand, is well
fitted by a population synthesis spectrum in which 50\% of the light at
4000~{\AA} is produced by 1.0$-$1.5 Gyr old stars. Allowing for the observed
dispersion in H$\delta$ equivalent widths, this would imply that about
$8.3(\pm$5)\% of the stars formed between 1 and 1.5 Gyr ago. The relative
fractions of normal and strong-lined galaxies then suggest that, on average,
E/S0 galaxies in clusters at $z=0.31$ formed less than about 1.5\% of their
stars in the last 2.5 Gyr and roughly $1.25(\pm$1)\% between 1.0 and 1.5 Gyr
ago. The mass fractions of young stars could therefore be similar for the all
the galaxies studied by Couch \& Sharples, the differences in H$\delta$
equivalent widths resulting from small differences in the ages of these stars.

Figure 4 summarizes the results of the previous paragraphs. For consistency, we
have reexpressed all the observational limits on the fraction of luminosity of
E/S0 galaxies produced by young stars as limits on the contribution in the $V$
band (Fig.~4{\it a}). The ages of the stars, which vary from one observation to
another, are indicated in billion years. Figures 4{\it b} and 4{\it c} show the
resulting limits on the mass fraction of stars formed in early-type galaxies as
a function of redshift. We have reexpressed constraints on the age of stellar
populations as constraints on their redshift of formation for two cosmological
models: a flat universe with $q_0=0.5$ and $h=0.45$ and an open universe with
$q_0=0.1$ and $h=0.55$ (where $h=H_0/ 100$~km~s$^{ -1}$~Mpc$^{ -1}$). Both
correspond to a present age of the universe of about 15 Gyr and lead to similar
predictions.  Figure 4 shows that the mass fraction of stars formed in E/S0
galaxies increases with redshift, from less than 1\% at $z\approx0$ to about
8\% at $z\approx1$. This result was anticipated earlier by Pickles (1989) on
the basis of his data. Moreover, the upper limits set by Schweizer \& Seitzer
(1992) on past rate of star formation in nearby field galaxies for their two
assumptions of Sb-Sb mergers and Sc-Sc mergers, obtained with an earlier
version of the population synthesis models used here (Charlot \& Bruzual 1991),
bracket the results for cluster galaxies in Figure 4. Thus, recent star
formation might be a generic feature of E/S0 galaxies both in clusters and in
the field.

We show two other limits in Figure 4. One is the lower limit on the mass
accretion rate by $L_{\star}$ galaxies in a standard model of hierarchical
clustering ({\it dashed line} in Fig.~4{\it b}). This is the minimum mass
fraction accreted by an $L_{\star}$ galaxy to a given redshift in a cold dark
matter-dominated universe (Carlberg 1990{\it a}, 1990{\it b}; T\'oth \&
Ostriker 1992). The steep rise of this limit at high redshift is not in
contradiction with our results because some of the accreted material might
already be in the form of old stars. In a low-density universe (Fig.~4{\it c}),
the crude limit on the mass accreted by an $L_{\star}$ galaxy obtained using
the scaling formulae of Carlberg (1990{\it b}) is much lower. T\'oth \&
Ostriker (1992) also derived an upper limit on the mass of material accreted
recently by spiral galaxies from the thinness and coldness of their disks.
Although spiral and elliptical galaxies are generally found in different
environments, it is interesting to see how this limit compares to that on the
mass of stars formed recently in E/S0 galaxies. According to T\'oth \&
Ostriker, a typical giant spiral galaxy must have accreted less than 4\% of its
disk mass in the last 5 Gyr. This is roughly 2\% of the mass of a typical,
$L_{\star}$, E/S0 galaxy, since elliptical galaxies have an average
mass-to-light ratio twice that of spiral galaxies (e.g., Faber \& Gallagher
1979). This upper limit therefore compares with that obtained by Rose (1985)
for E/S0 galaxies and is indicated by an open triangle in Figures 4{\it b} and
4{\it c}.

The evolution of the star formation rate in E/S0 galaxies shown in Figure 4
represents in reality an average over large redshift intervals, as the
horizontal error bars indicate. This result does not imply that E/S0 galaxies
should form stars at all times. In fact, the ages of intermediate-age stars
estimated from the absorption-line strengths in the spectra of the galaxies are
uncertain by a few billion years (Fig.~4{\it a}). As Figure 1 shows, when a
galaxy stops forming stars, the colors reach the values characteristic of old,
passively evolving stellar populations in less than 1 Gyr. Thus, although the
galaxies in a given cluster may present similar signatures of recent star
formation, there should be a dispersion in the ages and hence colors of
galaxies around the mean epoch of star formation predicted by Figure 4. The
identification by Couch \& Sharples (1987) of galaxies with different
Balmer-line strengths but associated with the same mass of new stars in
clusters at $z=0.31$ supports this expectation.

\centdbl{\bf 4. IMPLICATIONS FOR THE SPECTROPHOTOMETRIC}{\bf EVOLUTION AT HIGH
REDSHIFTS}

The star formation rates obtained in the previous section imply that E/S0
galaxies in clusters have undergone spectrophotometric evolution at redshifts
$0\simlt z\simlt 1.1$. We now evaluate this evolution using the population
synthesis models and compare our results with observations of high-redshift
clusters. In particular, Dressler \& Gunn (1983, 1990, 1992) have obtained $g$,
$r$, and $i$ photometry of 15 rich clusters at redshifts $0.04\leq z \leq0.94$.
They also measured the 4000 {\AA} breaks of bright ($r\simlt 22$~mag) galaxies
in each cluster. This spectroscopic subsample may be slightly incomplete but is
thought to include galaxies in the full range of colors in representative
proportions. More recently, Arag\'on-Salamanca et al. (1993) obtained $V$, $I$,
and $K$ photometry of 10 rich clusters at redshifts $0.55 \leq z \leq 0.92$,
two of which belong to the Dressler \& Gunn sample. They selected the galaxies
in the $K$ band, as opposed to the $r$-band selection by Dressler \& Gunn, to
reduce the chances of biasing the sample toward blue objects. However, unlike
Dressler \& Gunn, Arag\'on-Salamanca et al. do not have redshifts for their
galaxies; hence the most distant clusters may be significantly contaminated by
foreground objects. The clusters themselves in these samples were selected from
various catalogues, for which the reddest filter corresponds to the
photographic $N$ band ($\lambda_{\rm eff}\approx 8000$ {\AA}). The main results
of these studies are that the fraction of star-forming galaxies in clusters
increases with redshift (the Butcher-Oemler effect) and that beyond a redshift
of about 0.7, the reddest galaxies in clusters show evidence for evolution
consistent with passive evolution (the ``blueing of the red envelope"). In the
following, we examine how the evolution of the colors and 4000 {\AA} break
expected from recent star formation in E/S0 galaxies compares with the ranges
of colors and 4000 {\AA} breaks of galaxies in these observations. The
distributions of the spectrophotometric properties among galaxies in a cluster
will be addressed in the next section.

The history of star formation prior to the last major burst of star formation
in model galaxies is a free parameter, since it has little influence on the
subsequent evolution (Fig.~3). We recall that our main purpose in this paper is
to evaluate the minimum amount of evolution implied for E/S0 galaxies in
clusters by the signatures of recent star formation. We therefore consider a
conservative model, in which all E/S0 galaxies are assumed to initially be
elliptical galaxies formed at a redshift $z_{\rm F}=5$, and to then undergo
bursts of star formation at various redshifts at the rate estimated previously
from the low-redshift signatures of intermediate-age stars (we adopt throughout
this section $q_0=0.5$ and $h= 0.45$). We will also discuss later the case of
progenitor spiral galaxies and the effect of changing $z_{\rm F}$. The
evolution of the $V-K$ and $I-K$ colors of the model elliptical galaxy in the
absence of late bursts of star formation is shown by the dashed lines in Figure
5. The colors are zero-pointed to the {\it k}-corrected colors of a
non-evolving, bright elliptical galaxy (which has $V-K=3.25$ and $I-K=1.87$ at
$z\approx0$ for the filter response functions adopted in Fig.~5; Bruzual \&
Charlot 1993). As expected, the model elliptical galaxy is bluer at high
redshifts than a non-evolving elliptical galaxy but has $\Delta (V-K) \approx0$
and $\Delta(I-K)\approx0$ at $z\approx0$ (i.e., at an age of about 13.5 Gyr for
our adopted parameters). Also shown in Figure 5 are the observed colors of
galaxies in high-redshift clusters from the sample of Arag\'on-Salamanca et al.
(1993) and, in Figure 5{\it a}, the $V-K$ colors of galaxies in the Virgo and
Coma clusters from Bower et al. (1992). The Butcher-Oemler effect and the
blueing of the red envelope at high redshift are mostly apparent in the $V-K$
color, which is more sensitive than the $I-K$ color to variations of the
rest-frame blue light.

The solid lines in Figure 5 show the effects of new bursts of star formation on
the color evolution of the model elliptical galaxy. Again, this is the minimum
color evolution expected for E/S0 galaxies in clusters from their signatures of
intermediate-age stars at low redshifts. The different curves on the figure
correspond to the estimates at different redshifts of the mean mass fraction of
new stars in the progenitors of E/S0 galaxies from Figure 4{\it b}. As
expected, the bursts of star formation temporarily blue the galaxies, which
then rapidly recover the red colors characteristic of passively evolving
stellar populations (see Fig.~1). Figure 5 shows that the modest amount of star
formation predicted for E/S0 galaxies at high redshifts is sufficient to
explain the full range of blue colors seen in the data. We have checked that
the increase of the color range with redshift is primarily caused by the shift
of the filters into the rest-frame ultraviolet, which is more sensitive to the
presence of young stars than the optical and infrared light, rather than by the
increase of the star formation rate. The reason for this is that elliptical
galaxies have modest ultraviolet luminosities. Thus, for any amount of star
formation, blue massive stars dominate the ultraviolet spectrum. The strong
dependence of the color range on redshift is common to the models and data in
Figure 5. However, some of the observed blue objects may be foreground galaxies
(Arag\'on-Salamanca et al. 1993).

We compare in Figure 6{\it a} the 4000 {\AA} break evolution of the model
galaxies of Figure 5 with the observed 4000 {\AA} breaks of galaxies in
high-redshift clusters by Dressler \& Gunn (1990). For reference, we also show
the 4000 {\AA} breaks of field E/S0 galaxies from Hamilton (1985) and Spinrad
(1986). Again, the Butcher-Oemler effect and blueing of the red envelope are
manifest in the cluster data. In this case, however, the cluster memberships of
all galaxies are secure. At $z\approx0$, the model elliptical galaxy has the
standard \b4$\approx2.2$ of present-day E/S0 galaxies (Hamilton 1985; Spinrad
1986). Furthermore, the bursts of star formation at various redshifts produce
small 4000 {\AA} breaks that can match the observations. The range of \b4,
however, does not depend on redshift as sensitively as the range of colors
considered previously because the 4000 {\AA} break samples the rest frame
spectra of the objects. In fact, galaxies with 4000 {\AA} breaks of about 1.2
are found at all redshifts. Hence, if the signatures of intermediate-age stars
in low-redshift E/S0 galaxies are interpreted as recent bursts of star
formation superimposed on old elliptical galaxies, the bluest colors and
weakest 4000 {\AA} breaks observed for galaxies in high-redshift clusters can
be explained. However, the same range of spectrophotometric properties could be
obtained for much different amounts of star formation. This emphasizes again
the usefulness of constraints from absorption-line studies of low-redshift
cluster galaxies.

Several galaxies in Figures 5 and 6{\it a} appear to have observed colors that
are much redder, and 4000 {\AA} breaks that are much stronger, than the model
elliptical galaxy at $0.1\simlt z\simlt1$. This deserves comment because old,
passively evolving stellar populations are generally expected to generate the
reddest spectra. Extinction by dust is unlikely to be the explanation for these
red objects because the $I-K$ color appears to be at least as reddened as the
$V-K$ color. Furthermore, dust has little influence on the 4000 {\AA} break,
with \b4 increasing by less than 0.15 for an extinction $A_B=1$. Another
possible source of reddening is metallicity. The population synthesis models
used here are limited to solar metallicity. However, present-day elliptical
galaxies are observed to have an average metallicity of nearly twice solar with
a dispersion of about 50\% (Gonzalez 1993). We have evaluated the effect of
increasing and decreasing the metallicity by a factor of two in the models
using recent stellar evolutionary tracks and model atmospheres for non-solar
metallicities by Schaerer et al. (1993) and Kurucz (1992; see Charlot et al.
1993 for details). The results, indicated by the error bars in Figures 5 and 6,
show that most of the red galaxies can be understood as early-type galaxies
with less than twice solar metallicities. The few galaxies with very red colors
in Figure 5 could be background galaxies, even though similar objects do not
seem to appear in control field samples (Arag\'on-Salamanca et al. 1993). A
possible explanation may be that they underwent recently a burst of star
formation with an IMF deficient in low-mass stars (Charlot et al. 1993). This,
in fact, could also explain the very large 4000 {\AA} breaks of some cluster
galaxies in Figure 6{\it a}.

The predicted spectral evolution after the last major burst of star formation
is similar if cluster galaxies are assumed to be initially spiral instead of
elliptical galaxies. In Figure 6{\it b} we show the evolution of the 4000 {\AA}
break for models in which bursts of star formation are added to progenitor
spiral galaxies which formed stars continuously since $z_{\rm F}=5$. In this
case, \b4 remains nearly constant as long as star formation is maintained in
the underlying spiral galaxy ({\it dashed line}). Then, when a final burst
occurs and star formation stops, the 4000 {\AA} break increases rapidly as in
the case of a progenitor elliptical galaxy. In fact, as in Figure 6{\it a}, the
model galaxies in Figure 6{\it b} have at $z\approx0$ the typical
\b4$\approx2.2$ of present-day E/S0 galaxies. Hence, we conclude from Figures 5
and 6 that the signatures of intermediate-age stars in low-redshift E/S0
galaxies imply that these galaxies could have been some of the objects with the
bluest colors and weakest 4000 {\AA} breaks observed in high-redshift clusters.
This conclusion does not depend on the exact fraction of stars thought to have
formed recently in E/S0 galaxies, since similar ranges of colors and 4000 {\AA}
breaks could be obtained for different predictions. Finally, we note that
Figures 5 and 6 alone do not allows us to determine whether the progenitors of
E/S0 galaxies were preferentially spiral or elliptical galaxies in distant
clusters.

\centsh{\bf 5. DISCUSSION}

The main result of the previous sections is that a few billion years ago, the
progenitors of E/S0 galaxies in present-day clusters were at least temporarily
blued by modest amounts of star formation. Their colors and 4000 {\AA} breaks
were then indistinguishable from those of galaxies which make up the
Butcher-Oemler effect. We can use this result to set constraints on the global
evolution of galaxies in clusters in the following way. The conservative model
considered in \S4, in which the progenitors of all E/S0 galaxies are assumed to
be old elliptical galaxies undergoing late bursts of star formation, provides a
lower limit to the fraction of blue galaxies expected to be found in clusters
at high redshifts for two reasons: firstly, E/S0 galaxies could equally well be
the products of blue, spiral progenitor galaxies (Fig.~6{\it b}); and secondly,
E/S0 galaxies represent on average only 80$-$90\% of the galaxy population in
the cores of rich, concentrated nearby clusters. The remaining galaxies are
spiral and irregular galaxies. These have colors systematically redder in
clusters than in the field [$\Delta(B-V)\approx0.2$ mag], implying that many
early-type spiral galaxies in clusters have colors nearly as red as E/S0
galaxies ($B-V\approx 0.9$). However, the galaxies were presumably bluer in the
past. Hence, by modelling the evolution of only E/S0 galaxies and assuming that
these were old elliptical galaxies which underwent late bursts of star
formation, one obtains a lower limit on the fraction of blue galaxies in
high-redshift clusters. Any excess of blue galaxies in the observed color
distribution with respect to this prediction may then be identified as spiral
progenitors of E/S0 galaxies, or the progenitors of present-day spiral
galaxies, or even galaxies which have since faded away.

To investigate this further, we now compute the expected distribution in a
cluster at a redshift $z$ of the spectrophotometric property $C$ (for example,
a color) of progenitors of present-day E/S0 galaxies, under the conservative
assumption that these were initially elliptical galaxies formed at $z_{\rm
F}=5$. The mean mass fraction of stars added to the galaxies at $z$ can be
inferred from the observed trend in Figure 4{\it b}. This fraction, however,
could have been added at any redshift in an interval [$z+\sigma_+(z),
z-\sigma_-(z)$] corresponding to the uncertainties of a few billion years on
the ages of young stellar populations in low-redshift E/S0 galaxies (Fig.~4{\it
a}). Therefore, the distribution of property $C$ at a mean redshift \mz can be
expressed as
$$
N(C) \propto
\int_{z+\sigma_+(z)}^{z-\sigma_-(z)} dz' \sum t(C,z,z'),
\eqno(1)$$
where $\sum t(C,z,z')$ is the total time between $z+\sigma_+(z)$ and
$z-\sigma_-(z)$ during which a galaxy undergoing a burst at $z'$ is observed to
have the property $C$. Figure 4 indicates that $\sigma_+(z)$ and $\sigma_-(z)$
correspond typically to uncertainties on the age of about 1 Gyr at $z\simlt0.5$
and 2 Gyr at $z\simgt0.5$.

In the remainder of this paper, we identify the spectrophotometric property $C$
as the 4000 {\AA} break for three main reasons: (1) color distributions of
distant cluster galaxies are more difficult to interpret, since they generally
are subject to foreground contamination (e.g., Arag\'on-Salamanca et al. 1993);
(2) the 4000 {\AA} break has the advantage of sampling directly the rest-frame
spectra of galaxies, while the interpretation of observed colors involves {\it
k}-corrections, which depend on the unknown morphological types of the
galaxies; and (3) the 4000 {\AA} break is much less sensitive than colors to
the presence of dust in the galaxies (Charlot et al. 1993). Figure 7 shows the
expected distribution at different redshifts of the 4000 {\AA} break of
progenitors of low-redshift E/S0 galaxies calculated using the results of the
previous section and equation (1) above. We indicate for reference the expected
distribution of \b4 in the absence of late bursts of star formation, i.e., for
elliptical galaxies evolving passively since $z_{\rm F}=5$ (Fig.~7{\it a}). As
expected, the inclusion of new bursts of star formation produces a tail at
small \b4 in the histograms (Fig.~7{\it b}). The fraction of galaxies in the
tail increases with redshift, from about 12\% at $z=0.05$ to almost 30\% at
$z=0.5$. The reason for this is that the mass fraction of new stars, and hence
the time during which these dominate the spectra of the galaxies, increases
with redshift (Figs.~3{\it a} and 4{\it b}). The fraction of objects with 4000
{\AA} breaks smaller than passively evolving galaxies drops again under 20\% at
redshifts $z\simgt0.7$ because the model elliptical galaxy still has a small
4000 {\AA} break at these early ages. We note that shortening the timescale of
new bursts of star formation from our adopted value of 0.1 Gyr to 0.01 Gyr
would reduce the fraction of galaxies in the tail of the distribution by less
than 1\%.

In reality, the 4000 {\AA} break of early-type galaxies also correlates with
luminosity, an effect similar to the color-magnitude relation (\S3). Thus, the
distribution of 4000 {\AA} breaks for E/S0 galaxies in a cluster is expected to
be broadened by their spread in luminosity. The most natural way to compare our
predictions with observations would be to normalize the 4000 {\AA} breaks of a
sample of observed galaxies to a fixed absolute magnitude using the mean
\b4-magnitude relation in the sample (e.g., Butcher \& Oemler 1984).
Unfortunately, we cannot use this approach here because the magnitudes of
individual galaxies in the Dressler \& Gunn (1990) sample under consideration
are not available. Therefore, we must include in the models the influence of
the luminosity function of E/S0 galaxies on the distribution of 4000 {\AA}
breaks. We evaluate this effect using observations of galaxies in the Virgo
cluster. We adopt the gaussian approximation of the $B$-band luminosity
function of E/S0 galaxies in this cluster given by Sandage, Binggeli, \& Tamman
(1985). We may ignore dE/dS0 galaxies, since all galaxies in the Dressler \&
Gunn sample have luminosities larger than $0.1L_{\star}$ (Gunn 1989).
Furthermore, we estimate the \b4, $V$-magnitude relation for E/S0 galaxies in
Virgo using the recent determination of the $U-V$, $V$ color-magnitude relation
by Bower et al. (1992) and the tight correlation between \b4 and $U-V$ color
(Arag\'on-Salamanca et al. 1993). The \b4, $B$-magnitude relation can then be
evaluated using the $B-V$, $V$ color-magnitude relation (Sandage 1972). This is
the least certain step because the $B-V$, $V$ color-magnitude relation is not
very tight. Finally, we obtain the relative number of E/S0 galaxies with a
given 4000 {\AA} break from the \b4, $B$-magnitude relation and $B$ luminosity
function.

The distribution of \b4 obtained in this way for E/S0 galaxies in the Virgo
cluster is close to a gaussian distribution with a mean of 2.13 and a standard
deviation of about 0.15. For comparison, the mean and standard deviation of the
4000 {\AA} breaks of field galaxies observed at \mz$=0.05$ are 2.19 and 0.13,
respectively (Hamilton 1985; Spinrad 1986). Here, we are primarily interested
in the dispersion of the \b4 distribution, which we include in the models by
assuming that the distribution of 4000 {\AA} breaks of early-type galaxies has
the same shape at all redshifts as determined for E/S0 galaxies in the Virgo
cluster. The mean of the distribution, on the other hand, is taken to be the
value predicted by the models. In particular, at $z=0.05$ the model elliptical
galaxy considered above has \b4$=2.21$. We note that the similarity of the
color-magnitude relations in Virgo, Coma, and Abell~370 indicates that the
dispersion in \b4 for early-type galaxies may, in fact, not change
significantly out to at least $z\approx0.4$ (Arag\'on-Salamanca, Ellis, \&
Sharples 1991). Since the dispersion in \b4 only pertains to E/S0 galaxies, we
do not apply a correction to model galaxies which have just undergone a new
burst of star formation, so long as the amplitude of their 4000 {\AA} break is
more than 0.2 smaller than the expected value for a passively evolving galaxy
at the same redshift.

Figure 8{\it a} ({\it heavy histograms}) illustrates the broadening by the
\b4-magnitude relation of the predicted distributions of 4000 {\AA} breaks for
the progenitors of E/S0 galaxies from Figure 7{\it b}. Also shown are the
observed distributions of 4000 {\AA} breaks for galaxies in clusters at the
same redshifts from the sample of Dressler \& Gunn (1990; {\it dotted
histograms}); these include the spiral galaxy population. It appears that, as
expected, high-redshift clusters contain significantly more galaxies with small
4000 {\AA} breaks than predicted by our conservative model. To quantify this
result, we first estimate the characteristics of the dominant population of
early-type galaxies with large 4000 {\AA} breaks at the various redshifts in
both the models and data.  We use the ``biweight" central location and scale
estimators, which are particularly suited for non-gaussian distributions with
tails such as those investigated here (Beers, Flynn, \& Gebhardt 1990). The
resulting estimates of the central locations are shown by arrows above the
histograms in Figure 8{\it a}. At $z\simlt0.5$, the predicted and observed
distributions of \b4 for early-type galaxies have similar central locations.
However, the scale --- an estimate of width --- of the observed distributions
($\sim0.25$) is larger than that of the predicted distributions ($\sim0.15$).
This difference appears to result from an excess both of galaxies with
unusually large breaks and of galaxies with small breaks in the observed
distributions.

At $z\simgt0.7$, galaxies with large 4000 {\AA} breaks are not observed, and
the early-type population is no longer well defined. This is the blueing of the
red envelope discussed by Dressler \& Gunn (1990) and Arag\'on-Salamanca et al.
(1993). In the models, however, the distribution of early-type galaxies is
still well defined at these high redshifts, and the central location decreases
much more weakly. The absence of galaxies with strong 4000 {\AA} breaks may not
be common to all clusters at $z\simgt0.7$. In fact, this result has been
challenged by the recent discovery by Dickinson (1993) of several galaxies in
two clusters at $z\approx1.2$ with red $R-K$ colors consistent with those
predicted by our model elliptical galaxy. For reference, we also show in Figure
8{\it b} the results obtained when the progenitor elliptical galaxies of
present-day E/S0 galaxies are assumed to form at $z_{\rm F}=2$ instead of
$z_{\rm F}=5$. In this case, the central locations of the observed and
predicted histograms are in slightly better agreement at high redshift and
remain in good agreement at low redshift. The paucity of the observed sample at
\mz$=0.9$, however, does not allow us to favor a low $z_{\rm F}$.

We can now investigate the fraction of galaxies with 4000 {\AA} breaks
significantly smaller than E/S0 galaxies in the histograms of Figure 8. We
define here the fraction of galaxies with small 4000 {\AA} breaks in a cluster
to be the fraction of the total cluster population with \b4 at least 0.5
smaller than the central location of the model \b4 distribution. This appears
to be the minimum significant difference with respect to the 4000 {\AA} breaks
of early-type galaxies common to all model histograms in Figure 8 ({\it
triangles}). In this definition, we have used the central location of the
predicted distribution because it agrees with the observed one for $z\simlt0.4$
and is more clearly defined at higher redshift. The resulting fraction of
galaxies with small \b4 in Figure 8 is larger in the data than in the models at
all redshifts except for \mz$=0.5$. In fact, the observed sample at this
redshift includes the cluster 0016+16, which is known to be particularly
deficient in blue galaxies (Koo 1981; Butcher \& Oemler 1984). We then find
that, for $z_{\rm F} =5$, the predicted fractions of galaxies with small \b4
are 6\%, 4\%, 4\%, 3\%, and 3\%, and the observed ones are 10\%, 19\%, 4\%,
27\%, and 33\%, at \mz$= 0.05$, 0.4, 0.5, 0.7, and 0.9, respectively. The model
predictions are naturally lower than those obtained previously without
including the \b4-magnitude relation.

The above results taken at face value suggest that, under the most conservative
assumptions, the signatures of recent star formation in low-redshift E/S0
galaxies could explain at least 60\%, 21\%, 100\%, 11\%, and 9\% of the
observed population of star-forming galaxies in clusters at \mz$=0.05$, 0.4,
0.5, 0.7, and 0.9, respectively. The prediction for \mz$=0.5$ is probably not
representative of most clusters, since it is based on the unusual galaxy
population of 0016+16 (Koo 1981; Butcher \& Oemler 1984). Also, the prediction
for \mz$=0.05$ follows from only an upper limit by Rose (1985) on the mass
fraction of stars formed over the last 2.5 Gyr in nearby E/S0 galaxies, and
hence may be overestimated. Despite these uncertainties, there appears to be a
trend of increasing fraction of galaxies with small 4000 {\AA} breaks in
high-redshift clusters, which cannot be explained by the occurence of bursts of
star formation in old elliptical galaxies. According to the crude modelling
performed above, this unaccounted fraction would represent at least 40\% of the
observed star-forming galaxies at \mz$=0.05$, increasing to roughly 80\% at
\mz$=0.4$, and 90\% at \mz$=0.7$.

The most natural explanation for the excess population of galaxies with small
4000 {\AA} breaks in distant clusters is to invoke spiral galaxies. About
10$-$20\% of galaxies in the cores of rich, concentrated nearby clusters are of
late type (Butcher \& Oemler 1978). However, few of these galaxies are blue, as
many Sa and Sb galaxies have colors and 4000 {\AA} breaks comparable to those
of E/S0 galaxies (Dressler \& Schectman 1987; Oemler 1992). Hence, if only 4\%
of red spiral galaxies in present-day clusters were blue at \mz$=0.05$, and if
most of them (about 15\% of the total cluster population) were still blue at
\mz$=0.4$, the discrepancy between the predicted and observed fractions of
galaxies with small 4000 {\AA} breaks at $z\simlt0.4$ in Figure 8{\it a} could
be explained without having to invoke morphological evolution of spiral into
E/S0 galaxies (Fig.~6{\it b} shows that star-forming spiral galaxies at
$z\approx 0.4$ can have today a 4000 {\AA} break as strong as E/S0 galaxies).
In fact, recent spectral analyses and {\it HST} observations indicate that many
of the blue galaxies in clusters at $0.3\simlt z\simlt0.4$ are normal spiral or
``disky" galaxies (Couch et al. 1993; Dressler 1993; Jablonka \& Alloin 1993).
At a redshift of 0.7, however, the unaccounted fraction of galaxies with small
\b4 in Figure 8{\it a} is too high to be explained by the single color
evolution of spiral galaxies. A possible explanation would be for these excess
galaxies to be spiral progenitors of present-day E/S0 galaxies, implying that
morphological evolution has played a major role early on in clusters.
Alternatively, the excess star-forming galaxies at $z\simgt0.7$ could have
faded away by the present epoch (dynamical disruption, top-heavy IMF, etc.).

Hence, spectral analyses of low-redshift E/S0 galaxies and the 4000 {\AA} break
distributions of galaxies in distant clusters suggest that the Butcher-Oemler
effect originates from enhancement of the population of blue spiral galaxies at
high redshift. Out to a redshift $z\sim0.4$, the available data do not seem to
require morphological evolution and can be understood if the star formation
efficiency of spiral galaxies has declined substantially by the present epoch.
Several processes have already been proposed for causing such a decline (see
the review by Oemler 1992): ram-pressure stripping of the interstellar medium
of the galaxies through interactions with the hot intracluster medium;
galaxy-galaxy interactions and mergers; tidal interaction with the cluster
gravitational potential; and removal of the external gas supply of galaxies by
the dense cluster environment. At $z\simgt0.7$, on the other hand, more than
simple color evolution seems to be required to explain the large fraction of
galaxies with small 4000 {\AA} breaks in the clusters observed by Dressler \&
Gunn (1990). A natural explanation would be to invoke morphological evolution
of spiral into E/S0 galaxies, but the excess population of star-forming
galaxies could also have faded away by the present time. It appears, therefore,
that direct morphological studies at high redshifts will be the most secure way
to draw more definite conclusions about the evolution of galaxies in clusters.

We thank M. Dickinson, A. Dressler, and R. Ellis for several helpful
discussions. H.~Spinrad provided critical comments on the manuscript. This
research was supported at Berkeley in part by the NSF, and in part under the
auspices of the US Department of Energy at Lawrence Livermore National
Laboratory under contract number W-7405-Eng-48.

\vfill\eject

\par\vfill\eject
\centsh{\bf REFERENCES}

\Rf
Arag\'on-Salamanca, A., Ellis, R.S., \& Sharples, R.M. 1991, MNRAS, 248, 128
\Rf
Arag\'on-Salamanca, A., Ellis, R.S., Couch, W.J., \& Carter, E. 1993, MNRAS,
262, 764
\Rf
Beers, T.C., Flynn, K., \& Gebhardt, K. 1990, AJ, 100, 32
\Rf
Bower, R.G., Lucey, J.R., and Ellis, R.S. 1992, MNRAS, 254,601
\Rf
Bruzual A., G. 1983, ApJ, 273, 105
\Rf
Bruzual A., G., \& Charlot, S. 1993, ApJ, 405, 538
\Rf
Burstein, D., Faber, S.M., Gaskell, C.M., \& Krumm, N. 1984, ApJ, 287, 586
\Rf
Butcher, H., \& Oemler, A., Jr. 1978, ApJ, 219, 18
\Rf
Butcher, H., \& Oemler, A., Jr. 1984, ApJ, 285, 426
\Rf
Caldwell, N., Rose, J.A., Sharples, R.M., Ellis, R.S., \& Bower, R.G.
1993, AJ, 106, 473
\Rf
Carlberg, R.G. 1990{\it a}, ApJ, 350, 505
\Rf
---------. 1990{\it b}, ApJ, 359, L1
\Rf
Charlot, S., \& Bruzual A., G. 1991, ApJ, 367, 126
\Rf
Charlot, S., Ferrari, F., Mathews, G.J., \& Silk, J. 1993, ApJ, 419, L57
\Rf
Couch, W.J., Ellis, R.S., Sharples, R.M., \& Smail, I. 1993, in Observational
Cosmology, eds. G. Chincarini, A. Lovino, T. Maccacaro, \& D. Maccagni (San
Francisco: ASP), 240
\Rf
Couch, W.J., \& Sharples, R.M. 1987, MNRAS, 229, 423
\Rf
Dickinson, M. 1993, ApJ, in preparation
\Rf
Dressler, A. 1980, ApJ, 236, 351
\Rf
---------. 1993, in Observational Cosmology, eds. G. Chincarini, A. Lovino,
T. Maccacaro, \& D. Maccagni (San Francisco: ASP), 225
\Rf
Dressler, A., \& Gunn, J. 1982, ApJ, 263, 533
\Rf
---------. 1983, ApJ, 270, 7
\Rf
---------. 1990, in Evolution of the Universe of Galaxies, ed. R.G. Kron
(San Francisco: ASP), 200
\Rf
---------. 1992, ApJS, 78, 1
\Rf
Dressler, A., \& Schectman, S.A. 1987, AJ, 94, 899
\Rf
Evrard, A.E. 1991, MNRAS, 248, 8p
\Rf
Faber, S.M. 1972, A\&A, 20, 361
\Rf
---------. 1977, in The Evolution of Galaxies and Stellar Populations,
eds. B.M. Tinsley \& R.B. Larson (New Haven: Yale University Observatory),
157
\Rf
Faber, S.M., \& Gallagher, J.S. 1979, ARA\&A, 17, 135
\Rf
\Rf
Freedmann, W.L. 1992, AJ, 104, 1349
\Rf
Gonzalez, J.J. 1993, PhD thesis, Univ. of California at Santa-Cruz
\Rf
Gunn, J.E. 1989, The Epoch of Galaxy Formation, eds.  C.S. Frenk, R.S. Ellis,
T. Shanks, A.F. Heavens, \& J.A. Peacock (Dordrecht: Kluwer), 167
\Rf
Hamilton, D. 1985, ApJ, 297, 371
\Rf
Jablonka, P., \& Alloin, D., 1993, A\&A, submitted
\Rf
Koo, D.C. 1981, ApJ, 251, L75
\Rf
Kurucz, R.L. 1992, in IAU Symposium 149, The Stellar Populations of Galaxies,
eds. B. Barbuy \& A. Renzini (Dordrecht: Reidel), 225
\Rf
Lauer, T.R., 1988, ApJ, 325, 49
\Rf
Lavery, R.J., \& Henry, J.P. 1988, ApJ, 330, 596
\Rf
Mellier, Y., Soucail, G., Fort, B., \& Mathez, G. 1988, A\&A, 199, 13
\Rf
O'Connell, R.W. 1980, ApJ, 236, 430
\Rf
---------. 1986, in Spectral Evolution of Galaxies, eds.
C. Chiosi \& A. Renzini (Dordrecht: Reidel), 321
\Rf
Oemler, A., Jr. 1992, in Clusters and Superclusters of Galaxies,
ed. A.C. Fabian (Dordrecht: Kluwer), 29
\Rf
Pickles, A.J. 1985, ApJ, 296, 340
\Rf
---------. 1989, The Epoch of Galaxy Formation, eds.
C.S. Frenk, R.S. Ellis, T. Shanks, A.F. Heavens, \& J.A. Peacock
(Dordrecht: Kluwer), 191
\Rf
Pickles, A.J., \& van der Kruit, P.C. 1988, in Toward Understanding
Galaxies at Large Redshift, eds. R.G. Kron \& A. Renzini
\Rf
Pritchet, C. 1977, ApJS, 35, 397
\Rf
Rose, J. 1984, AJ, 89, 1258
\Rf
---------. 1985, AJ, 90, 1927
\Rf
Sandage, A. 1972, ApJ, 176, 21
\Rf
Sandage, A., Binggeli, B., \& Tammann, G.A. 1985, AJ, 90, 1759
\Rf
Schaerer, D., Meynet, G., Maeder, A., \& Schaller, G. 1993, A\&AS, 98, 523
\Rf
Schaller, G., Schaerer, D., Meynet, G., \& Maeder, A. 1992, A\&AS, 96, 269
\Rf
Schweizer, F., \& Seitzer, P. 1992, AJ, 104, 1039
\Rf
Spinrad, H. 1986, PASP, 98, 269
\Rf
Spinrad, H., \& Taylor, B.J. 1969, ApJ, 157, 1279
\Rf
T\'oth, G., \& Ostriker, J.P. 1992, ApJ, 389, 5
\Rf
Worthey, G. 1993, ApJ, 409, 530
\Rf
Worthey, G., Faber, S.M., \& Gonzalez, J.J. 1992, ApJ, 398, 69

\vfill\eject
\centsh{\bf FIGURE CAPTIONS}

\noindent {FIG. 1.} --- Evolution of the $U-V$, $V-I$, and $I-K$ colors and
4000 {\AA} breaks of two
model galaxies undergoing a last burst of star formation
which involves 10\% of their final mass at an age of 6 Gyr. The solid line
corresponds to a 0.5 Gyr burst stellar population (model elliptical galaxy),
and the dashed line to a stellar population with constant star formation rate
for the first 6 Gyr (model spiral galaxy). The models have solar metallicity
and the Salpeter IMF from 0.1 to 100 $M_{\odot}$. The filter response functions
correspond to the RGO CCD $U$ (which includes the atmosphere), $V$, and
Kron-Cousins $I$ filters and the UKIRT IRCAM $K$ filter.
\vskip8pt

\noindent {FIG. 2.} --- Evolution of the H$\delta$ absorption equivalent widths
and \caidx indices as a function of  $U-V$ and $V-K$ colors for the model
galaxies of Fig.~1. The evolution is shown from an age of 6 Gyr, when the
bursts
occur ({\it triangles}). The tickmarks indicate the positions of the models at
successive steps of 0.25 Gyr, until the age of 8 Gyr. The filled triangle and
solid
line correspond to the case of an underlying elliptical galaxy, and the open
triangle and dashed line to that of an underlying spiral galaxy. The filter
response functions are the same as in Fig.~1.
\vskip8pt

\noindent {FIG. 3.} ---  Evolution of the fractional contribution to the
$V$ luminosity by stars produced in a last burst of star formation in a
galaxy: ({\it a}) for a burst involving 10\% ({\it solid line}) and 2\%
({\it dashed line}) of the final mass, occuring in a 6 Gyr old elliptical
galaxy; ({\it b}) for a burst involving 10\% of the final mass, occuring in a
3, 6, and 9 Gyr old elliptical galaxy; and ({\it c}) for a burst involving
10\% of the final mass, occuring in a 6 Gyr old  elliptical ({\it solid line})
and spiral ({\it dashed line}) galaxies.
\vskip8pt

\noindent {FIG. 4.} --- ({\it a}) Observed contributions by intermediate-age
stars to the $V$ luminosity of E/S0 galaxies in low-redshift clusters inferred
from stellar absorption-line studies. The corresponding ranges of stellar
ages are indicated next to the sources (see text for details). ({\it b}) Mass
fraction of new stars in the progenitors of E/S0 galaxies as a function
of redshift derived from ({\it a}) for the cosmology $h=0.45$ and $q_0=0.5$.
The solid lines indicate the upper limits on the past star formation rate
in nearby E/S0 field galaxies set by the merger models of Schweizer \& Seitzer
(1992), for Sb-Sb type ({\it lower curve}) and Sc-Sc type ({\it upper curve})
mergers. The dashed line is a lower limit on the mass accretion rate by an
$L_\star$ galaxy in a standard model of hierarchical clustering (Carlberg
1990{\it a}, 1990{\it b}). The open triangle is a comparative upper limit
on the mass recently accreted by spiral galaxies from T\'oth \& Ostriker (1992;
see \S3 of the present paper). The horizontal error bars correspond to the
uncertainties on the ages of intermediate-age stars detected in low-redshift
E/S0 galaxies. The vertical error bars represent the uncertainties on the
observed fraction of $V$ light contributed by these stars and the uncertainties
on the determination of their contribution to the mass for the allowed ranges
of ages. ({\it c}) Same as ({\it b}) for the case $h=0.55$ and $q_0=0.1$.

\vskip8pt

\noindent {FIG. 5.} --- Evolution of the ({\it a}) $V-K$ and ({\it b}) $I-K$
colors of model galaxies forming at $z_{\rm F}=5$ as a function of
redshift for $h=0.45$ and $q_0=0.5$. The colors are zero-pointed to the {\it
k}-corrected colors of a non-evolving, present-day elliptical galaxy. The
thick dashed line corresponds to a model elliptical galaxy, and the solid
lines to the effect of adding bursts of star formation to this model with
the mass fractions of new stars taken from Fig.~4{\it b}. The data are the
observed colors of galaxies in high-redshift clusters from Arag\'on-Salamanca
et al. (1993). The associated uncertainties are typically less than 0.2
mag in both the $V-K$ and $I-K$ colors. The error bars on the figure indicate
the
effect on the predicted colors of increasing and decreasing the metallicity by
a
factor of two in the models. The filter response functions are the same as in
Fig.~1.
\vskip8pt

\noindent {FIG. 6.} --- ({\it a}) Evolution of the 4000 {\AA} break for the
models galaxies of Fig.~5. The data are the observed 4000 {\AA} breaks of
galaxies in high-redshift clusters from Dressler \& Gunn (1990; {\it filled
circles}) and of field galaxies from Hamilton (1985; {\it triangles}: program
galaxies; and {\it stars}: serendipitous galaxies) and Spinrad (1986; {\it
squares}). The observational uncertainties are of the order of 5\%.
({\it b}) Same as ({\it a}) but in the case where the progenitor galaxies
are assumed to be spiral instead of elliptical galaxies. The error bars on
the figure indicate the effect on the predicted \b4 of increasing and
decreasing the metallicity by a factor of two in the models.
\vskip8pt

\noindent {FIG. 7.} --- Model distributions of the 4000 {\AA} break for
progenitors of E/S0 galaxies in clusters at different redshifts ({\it top to
bottom}) computed using eq.~(1); ({\it a}) for passively evolving elliptical
galaxies forming at $z_{\rm F}=5$; and ({\it b}) when including bursts of star
formation at the rate predicted by the results of Fig.~4{\it b}.
\vskip8pt

\noindent {FIG. 8.} --- ({\it a}) The heavy histograms correspond to the
broadening by the \b4-magnitude relation of the model histograms of Fig.~7{\it
b} (see text). The dotted histograms show the observed distributions of 4000
{\AA} breaks of galaxies in clusters at the different redshifts (from Dressler
\& Gunn 1990). In each panel, the number of galaxies defining the observed
histogram is indicated. The down arrows show the central locations of the
model and observed distributions calculated using the biweight estimator (see
text). The open triangle indicates the value of the 4000 {\AA} break 0.5 away
from the central location of the model distribution. This is the limit under
which a galaxy is classified as having a small 4000 {\AA} break. ({\it b})
Same as ({\it a}) for the case $z_{\rm F} =2$.
\vskip8pt

\vfill\eject\end